\documentstyle[12pt]{article}

\def \beq {\begin{equation}}
\def \eeq {\end{equation}}

\begin{document}

\begin{titlepage}

\begin{flushright}
IFUSP/P-1112 \\
gr-qc@9405054
\end{flushright}

\begin{center}
{\Large{\bf Is minimal coupling procedure compatible with minimal action
principle?}}\footnote{This Essay was awarded a
Honorable Mention for 1994 by the Gravity Research Foundation.}

\vspace{1cm}

{\large{\bf Alberto Saa}}\footnote{Supported by FAPESP.}

\vspace{.5cm}

{\it Instituto de F\'{\i}sica \\
Universidade de S\~ao Paulo, Caixa Postal 20516 \\
01452-990 S\~ao Paulo,  Brazil\\}

\vspace{1cm}

{\bf Abstract}
\end{center}
{\small
When space-time is assumed to be non-Riemannian the minimal coupling
procedure (MCP) is not compatible, in general, with minimal action
principle (MAP). This means that the equations gotten by applying MCP to the
Euler-Lagrange equations of a Lagrangian $\cal L$ do not coincide with
the Euler-Lagrange equations of the Lagrangian obtained by applying MCP
to $\cal L$. Such compatibility can be restored if the space-time admits
a connection-compatible volume element. We show how these concepts can alter
qualitatively the predictions of the Einstein-Cartan theory of gravity.}

\vspace{2cm}

\end{titlepage}

\newpage

Minimal coupling procedure (MCP) provides us with an useful rule to get
the equations for any physical field on non-Minkowskian manifolds starting
from their versions of Special Relativity (SR). When studying classical
fields on a non-Minkowskian manifold $\cal X$ we usually require that the
equations of motion for such fields have an appropriate SR limit. There are,
of course, infinitely many covariant equations on $\cal X$ with the same
SR limit, and MCP solves this arbitrariness by saying that the relevant
equations should be the ``simplest'' ones. MCP can be heuristically formulated
as follows. Considering the equations of motion for a classical field in
the SR, one can get their version for a non-Minkowskian space-time $\cal X$
by changing the partial derivatives by the $\cal X$ covariant ones and the
Minkowski metric tensor by the $\cal X$ one. MCP is also
used for the quantum
analysis of gauge fields, where the gauge field is to be interpreted
as a connection, and it is in spectacular agreement with
experience for QED.

Suppose now that the SR equations of motion for a classical field follow
from an action functional via minimal action principle (MAP). It is natural to
expect that the equations obtained by using MCP to the SR equations
coincide with the Euler-Lagrange equations of the action obtained
via MCP of the SR one. This can be better visualized with the help of the
following diagram
\setlength{\unitlength}{1mm}
$$
\addtocounter{equation}{1}
\newlabel{diagr}{{1}{1}}
\hspace{106pt}
\begin{picture}(52,28)
\put(3,20) {$ {\cal C}_{ {\cal L}_{\rm SR} }$}
\put(7,18){\vector(0,-1){9}}
\put(3,5){$ E({\cal L}_{\rm SR}) $}
\put(45,20){${ \cal C_{L_X} }$ }
\put(40,5){$ E({\cal L}_{\cal X})$}
\put(47,18){\vector(0,-1){9}}
\put(12,22){\vector(1,0){30}}
\put(17,7){\vector(1,0){22}}
\put(24,24){${\scriptstyle \rm MCP}$}
\put(27,9){${\scriptstyle \rm MCP}$}
 \put(8,13){${\scriptstyle \rm MAP}$}
\put(48,13){${\scriptstyle \rm MAP}$}
\end{picture}
\hspace{116pt}\raise 7ex \hbox{(\theequation)}
$$
where $E({\cal L})$ stands to the Euler-Lagrange equations for the
Lagrangian $\cal L$, and ${\cal C}_{\cal L}$ is the equivalence class
of Lagrangians, ${\cal L}'$ being equivalent to $\cal L$ if
$E({\cal L}')=E({\cal L})$. The diagram (\ref{diagr}) is verified when
MCP is used for gauge fields and for General Relativity. We say that MCP is
compatible with MAP if (\ref{diagr}) holds. We stress that if (\ref{diagr})
does not hold we have another arbitrariness to solve, one needs to choose
one between two equations, as we will shown with a simple example.

It is not difficulty to check that MCP is not compatible with MAP,
in general, when space-time is assumed to be non-Riemannian, as for example
in the Einstein-Cartan theory of gravity\cite{hehl},
where the linear connection
$\Gamma^\alpha_{\mu\nu}$ is not symmetrical in its lower indices, but is
metric-compatible, $D_\alpha g_{\mu\nu}=0$. Let us examine for simplicity
the case of a massless scalar field $\varphi$ in the frame of Einstein-Cartan
gravity\cite{saa1}. The equation for $\varphi$ in SR is
\beq
\partial_\mu\partial^\mu\varphi=0,
\label{e2}
\eeq
which follows from the minimization of the action
\beq
\label{act}
S_{\rm SR} =
\int d{\rm vol}\, \eta^{\mu\nu}\partial_\mu\varphi\partial_\nu\varphi.
\eeq
Using MCP to (\ref{act}) one gets
\beq
\label{act1}
S_{\cal X} = \int d{\rm vol}\, g^{\mu\nu}
\partial_\mu\varphi\partial_\nu\varphi,
\eeq
and using the canonical volume element for $\cal X$, $
d{\rm vol} = \sqrt{g}d^nx$, we get the following equation from the
minimization of (\ref{act1})
\beq
\label{aa22}
\frac{1}{\sqrt{g}}\partial_\mu \sqrt{g}\partial^\mu\varphi = 0.
\eeq
It is clear that (\ref{aa22}) does not coincide in general with the
equation obtained via MCP of (\ref{e2})
\beq
\label{e3}
\partial_\mu\partial^\mu\varphi + \Gamma^\mu_{\mu\alpha}
\partial^\alpha\varphi =
\frac{1}{\sqrt{g}}\partial_\mu \sqrt{g}\partial^\mu\varphi
+ 2 \Gamma^\mu_{[\mu\alpha]} \partial^\alpha\varphi = 0.
\eeq
We have here an ambiguity, the equations (\ref{aa22}) and (\ref{e3}) are in
principle equally acceptable ones, to choose one of them corresponds to choose
as more fundamental the equations of motion or the action formulation from
MCP point of view. As it was already said, we do not have such ambiguity
when MCP is used to gauge fields and when space-time is assumed to be
a Riemannian manifold. This
is not a feature of massless scalar fields, all matter fields have the
same behaviour in the frame of Einstein-Cartan gravity. The incompatibility
of MCP and MAP for fermionic fields in the Einstein-Cartan gravity is
well known\cite{venzo}.

An accurate analysis of the diagram (\ref{diagr}) reveals that the source
of the problems of compatibility between MCP and MAP is the volume element
of $\cal X$\cite{saa2}.
It turns out that if $\cal X$ admits a connection-compatible
volume element, the diagram (\ref{diagr}) holds for all matter fields. A
connection-compatible volume element $d{\rm vol}=j(x)d^nx$ is such
that
\beq
D_\alpha j(x) = 0.
\eeq
It is easy to check that the canonical volume element
$d{\rm vol}=\sqrt{g}d^nx $ is compatible with the connection for a
Riemannian manifold and that it is not for a Riemann-Cartan manifold,
the space-time of the Einstein-Cartan gravity. A Riemann-Cartan manifold
admits a connection-compatible volume element if\cite{saa2}
\beq
\label{cond}
\Gamma^\mu_{[\mu\alpha]} = \partial_\alpha\Theta(x),
\eeq
in this case the connection-compatible volume element is
$d{\rm vol} = e^{2\Theta}\sqrt{g}d^nx$.

It is not usual to find in the literature applications where volume
elements different from the canonical one are used. In our case the
new volume element appears naturally, in the same way that we require
compatibility conditions between the metric tensor and the linear
connection we can do it for the connection and volume element. If
we remember that in a manifold with torsion there are no infinitesimal
parallelograms  and so there are no infinitesimal parallel cubes, there
are no reasons a priori to expect that the notion of volume of Riemannian
geometry be preserved in the presence of torsion. It is also important
to stress that any volume element that differs from the canonical one
by the multiplication of a positive function defines, in principle,
an acceptable notion of volume\cite{berger}.

With the use of the connection-compatible volume element in the
action formulation for Einstein-Cartan gravity we can have qualitatively
different predictions. The scalar of curvature for a Riemann-Cartan
manifold, present in the Hilbert-Einstein action, is given by the
Riemannian scalar of curvature plus terms quadratic in the torsion.
Due to (\ref{cond}) such quadratic terms will provide a differential
equation for $\Theta$, what will allow for non-vanishing torsion
solutions for the vacuum. Torsion can propagate if the space-time
admits a connection-compatible volume element, the torsion mediated
interactions loose their contact aspect. As to the matter fields, the
use of the connection-compatible volume element, besides of guarantee
that the diagram (\ref{diagr}) holds, brings also qualitative changes.
For example, it is possible to have a minimal interaction between
Maxwell fields and torsion preserving gauge symmetry. Another point
of interest is that the peculiar $\Theta$-dependance of the
connection-compatible volume element shows that such notion of volume
can have relevance to the study of dilaton gravity.

It is important to note that the restriction that space-time should
admit a connection-compatible volume element arises even in the case where
MAP is not used. It appears as integrability condition for the Maxwell
equations obtained by using MCP to the SR ones in the differential forms
formulation\cite{saa3}.

\end{document}